\documentclass[preprint]{aastex}

\def\msol{M$_{\odot}$}

\def\Pa{P${\alpha}$}

\def\Bg{Br${\gamma}$}

\def\micro{$\rm \,\mu m$}
\def\arcsec{\,''}

\def\deg{$^\circ$}
\begin{document}

\title{Massive Clusters in the Inner Regions of NGC 1365: Cluster Formation and
Gas Dynamics in Galactic Bars}

\author{Bruce G. Elmegreen} \affil{IBM Research Division, T.J. Watson
  Research Center, 1101 Kitchawan Road, Yorktown Heights, NY 10598, USA}
\email{bge@watson.ibm.com}

\author{Emmanuel Galliano}
\affil{Observatorio Nacional, Rua
   General Jos\'e  Cristino, 77, 20921-400, S\~ao
   Cristov\~ao,  Rio  de  Janeiro,  Brazil}
\email{egallian@on.br}

\author{Danielle Alloin} \affil{Laboratoire  AIM,
   CEA/DSM-CNRS-Universit\'e  Paris  Diderot,  IRFU/Service
  d'Astrophysique,  B\^at.709,  CEA/Saclay,  F-91191  Gif-sur-Yvette
   Cedex,  France} \email{danielle.alloin@cea.fr}

\begin{abstract}
Cluster formation and gas dynamics in the central regions of barred
galaxies are not well understood. This paper reviews the environment of
three $10^7\,M_\odot$ clusters near the inner Lindblad resonance of the
barred spiral NGC 1365. The morphology, mass, and flow of HI and CO gas
in the spiral and barred regions are examined for evidence of the
location and mechanism of cluster formation. The accretion rate is
compared with the star formation rate to infer the lifetime of the
starburst. The gas appears to move from inside corotation in the spiral
region to looping filaments in the interbar region at a rate of
$\sim6\;M_\odot$ yr$^{-1}$ before impacting the bar dustlane somewhere
along its length. The gas in this dustlane moves inward, growing in
flux as a result of the accretion to $\sim40\;M_\odot$ yr$^{-1}$ near
the ILR. This inner rate exceeds the current nuclear star formation
rate by a factor of 4, suggesting continued buildup of nuclear mass for
another $\sim0.5$ Gyr. The bar may be only 1-2 Gyr old. Extrapolating
the bar flow back in time, we infer that the clusters formed in the bar
dustlane outside the central dust ring at a position where an interbar
filament currently impacts the lane. The ram pressure from this impact
is comparable to the pressure in the bar dustlane, and both are
comparable to the pressure in the massive clusters.  Impact triggering
is suggested. The isothermal assumption in numerical simulations seems
inappropriate for the rarefraction parts of spiral and bar gas flows.
The clusters have enough lower-mass counterparts to suggest they are
part of a normal power law mass distribution. Gas trapping in the most
massive clusters could explain their [NeII] emission, which is not
evident from the lower-mass clusters nearby.
\end{abstract}


\section{Introduction}
Massive clusters in the inner regions of barred spiral galaxies are
often observed as ``hotspots'' in the Lindblad resonance ring (Morgan
1958; Sersic \& Pastoriza 1965). They have been studied using visible,
ultraviolet, near-infrared and radio wavelengths (e.g. Hummel et al.
1987; Benedict et al. 1993; Barth et al. 1995; Meurer et al. 1995;
Tacconi-Garman et al. 1996; Maoz et al. 1996; B\"oker et al. 2008). In
a recent study using near-infrared (NIR) and mid-infrared (MIR) images
and spectra, Galliano et al. (2005, 2008; hereafter G08) found three
compact clusters with masses of around $10^7\; M_\odot$ in the ILR ring
region of NGC 1365 (see also Kristen et al. 1997). These clusters are
somewhat equally spaced in a dense dusty region where their extinction
and MIR continuum emission suggest they are still embedded. Their ages
are between 6 and 8 Myr.

This paper reviews the environment of the three clusters in an attempt
to understand how they formed. We consider the cluster ages and masses
from G08, the distribution of gas mass and velocity from HI
observations by J\"ors\"ater \& van Moorsel (1995; hereafter JM95) and
CO observations by Sakamoto et al. (2007; hereafter S07), the likely
gas flow in the bar using models of NGC 1365 by Lindblad, Lindblad \&
Athanassoula (1996, hereafter L96), and the total accretion rate in
comparison to the star formation rate. These observations, along with a
detailed optical image of dust filaments in the galaxy, suggest that
the gas inside corotation spirals into the interbar region and hits the
bar dustlane after half of a rotation relative to the pattern. There it
shocks against the gas already in the bar dustlane, and both fall
directly to the central region. The result is a high pressure at
various places along the dustlane that can drive massive cluster
formation, and a high gas accretion rate to the center that can sustain
the observed starburst for several hundred Myr.

The flow of gas to the inner regions of barred galaxies is well
observed (e.g. Ishizuki et al. 1990; Regan, Vogel \& Teuben 1997;
Regan, Sheth, \& Vogel 1999; Knapen et al. 2000), as is the
accumulation of gas after this inflow has occurred (e.g., Kenney et al.
1992; Sakamoto et al. 1999; Sheth et al. 2005; Jogee et al. 2005). The
gas often makes a ring near the ILR (e.g., Regan et al. 2002) as a
result of gas shocking where the stable orbits change from aligned with
the bar to perpendicular (Sanders \& Huntley 1976; Combes \& Gerin
1985; Regan \& Teuben 2003). Star formation in or near this ILR ring is
common (see review in Buta \& Combes 1996; Kormendy \& Kennicutt 2004).
There have been several models for this star formation, including
gravitational instabilities in the ILR ring (Elmegreen 1994),
gravitational instabilities in dense spurs preceding the straight dust
lane (Sheth et al. 2000, 2002; Asif et al. 2005; Zurita \& P\'erez
2008), and gas impact triggering where the ring meets the straight bar
dust lane (B\"oker et al. 2008; Meier, Turner, \& Hurt 2008). Aspects
of all three models are evident in the present observations. Zurita et
al. (2004) noticed, in addition, an enhancement in star formation at
minima of the velocity gradient along the bar dust lane in NGC 1530;
they suggested that local shocks and compressions at these gradient
minima trigger star formation.

We also consider the unusually large cluster masses and question
whether they are part of a power law mass function, as typically found
in galaxy disks, or a characteristic mass in some physical process that
produces a peaked mass function. The difference in mass functions is
important for old globular clusters, which have been claimed in various
studies to have evolved from either one or the other of these two
functions (e.g., compare Vesperini 2000, Fall \& Zhang 2002, Parmentier
\& Gilmore 2005). We see tentative evidence for lower-mass clusters in
the dustlane, suggesting there is an underlying power law in the
cluster mass function. Maoz et al. (2001) found $M^{-2}$ power law mass
functions for ILR ring clusters in two other galaxies.

In what follows, section \ref{sect:env} provides a summary of the
physical parameters of the three cluster environments, while in
section~\ref{sect:where}, we discuss their likely formation site. In
sections~\ref{sect:flow} and~\ref{sect:accretion}, we examine the gas
accretion, star formation rates, and implications for the bar age.
Information about the cluster mass function is examined in
section~\ref{sect:cmf}. The removal of gas from massive clusters is
discussed in section~\ref{sect:remove}, and a possible mechanism for
their formation is in section~\ref{sect:formation}. The conclusions are
in section~\ref{sect:concl}.

\section{Characteristics and environment of the three ILR clusters in NGC
1365}\label{sect:env}

The three massive clusters in NGC 1365, designated M4, M5, and M6, were
discovered in the MIR by Galliano et al. (2005) and studied in more
detail by G08. They were also observed as radio continuum sources by
Sandqvist et al. (1995), Forbes \& Norris (1998), and Morganti et al.
(1999), and detected in CO(2-1) by S07. In what follows, we review the
properties of the three clusters given by G08 using a distance to NGC
1365 of 18.6 Mpc. The CO properties are summarized from S07, with
values converted from the distance they assume, 17.95 Mpc, to 18.6 Mpc.

Figure \ref{schematic_version2} shows an image of NGC 1365 from ESO. It
is a combination from three exposures with FORS1: B(blue), V(green),
and R(Red)
\footnote{http://www.eso.org/public/outreach/press-rel/pr-1999/phot-08-99.html}
with illustrations of various topics discussed in this paper. The three
clusters are associated with dust structures at the position where the
north-eastern bar dustlane enters the ILR dust ring. Cluster M5 is at
the edge of the dustlane and more prominent optically. This is
consistent with extinctions from the Br$\gamma$/Br$\alpha$ ratio, which
equal $A_V=13.5,$ 3.2, and 8.5 mag., respectively for M4, M5, and M6
(G08).  Bright nebular emission from P$\alpha$, Br$\gamma$, Br$\alpha$,
and [NeII] suggest local ionization of the dustlane by the clusters;
2$\mu m$ H$_2$ lines indicate local heating of molecules. Local
ionization is also suggested by an unusually strong $8.6\mu m$ feature
that is presumably from ionized PAHs. Thus the clusters are most likely
embedded in dust and associated with dense gas (G08).

There is an unusual lack of [ArIII] and [SIV] emission from all three
clusters which indicates that the radiation field is weak in the
far-ultraviolet (G08). This implies that the most massive remaining
stars are only $20-25 \;M_\odot$, and therefore suggests a time larger
than 6 Myr from the last event of significant star formation. The
presence of non-thermal centimeter emission and 2.3\micro~CO absorption
bands also constrains the age to be greater than several Myr. An upper
limit to the cluster ages is $\sim8$ Myr from the requirement that the
total mass in all three clusters be significantly less than the gas
mass in the inner kpc of the galaxy. For an age of 7 Myr, the cluster
masses are on the order of $10^7$\msol.

The ILR region of NGC 1365 was mapped at 2\arcsec~resolution in three
isotopes of CO(2-1) by S07. They assumed a conversion of $^{12}$CO to
H$_2$ equal to $X_{\rm CO}=0.5\times10^{20}$ cm$^{-2}$ $(\rm
{km\;s^{-1}})^{-1}$, which is comparable to that estimated for the
center of the Milky Way but smaller than the usual value applied to
galactic disks. High temperature CO would make $X_{\rm CO}$ small like
this. They checked $X_{\rm CO}$ by comparing the $^{12}$CO(2-1) mass to
the C$^{18}$O(2-1) mass and got the reasonable result that the latter
was smaller by a factor of $\sim2$, which is to be expected for the
denser C$^{18}$O(2-1) regions compared to the total. We assume the same
conversion factor here.  A similar low CO conversion factor was
proposed by Meier, Turner, \& Hurt (2008) for nuclear emission in
Maffei 2.

The total CO mass in the inner one kpc radius was determined by S07 to
be $ 9\times10^8\;M_\odot$, which converts to $9.7\times10^8\;M_\odot$
at the higher distance. (In what follows, we keep additional
significant figures like this in evaluations from other studies, even
though the observations and conversion factors may not warrant this
much accuracy, to keep track of the various distance conventions. The
final results are rounded off in summary statements.) The average
column density is $ 290\;M_\odot\,{\rm pc}^{-2}$. The region enclosing
clusters M4, M5, and M6 has a higher mean column density, $
\sim500\;M_\odot\,{\rm pc}^{-2}$, peaking at around $
800\;M_\odot\,{\rm pc}^{-2}$. Cluster M4 has a distinct molecular cloud
associated with it (S07), with a distance-converted mass of $
5.4\times10^7\;M_\odot$ and a peak column density at the limit of
resolution of $ \sim900\;M_\odot\,{\rm pc}^{-2}$. This cloud mass is
$\sim5.4$ times the mass of the cluster so the efficiency of star
formation was $\sim1/6.4=16$\%. Clusters M5 and M6 also have associated
CO peaks (S07), although they are not as prominent as the one around
M4.

The CO mass in the entire bar was estimated by Sandqvist et al. (1995)
to be $ 20\times10^9\;M_\odot$ using $X_{\rm CO}=2.3\times10^{20}$
cm$^{-2}$ $(\rm {km\;s^{-1}})^{-1}$.  They assumed a distance of 20
Mpc. With our distance of 18.6 Mpc, the bar CO mass becomes
$17\times10^9\;M_\odot$, and with the Sakamoto et al. (2007) value of
$X_{\rm CO}$, the bar CO mass is $ 3.8\times10^9\;M_\odot$. According
to the previous paragraph, $ 0.97\times10^9\;M_\odot$, or 26\% of the
total bar CO is currently in the central 1 kpc radius.  There is little
HI emission in the bar region (Sandqvist et al 1995), so this total is
what is available for ILR star formation after it accretes to the
center.

The velocity dispersion of the gas that formed the clusters can be
estimated from the cluster virial speed, because a cluster forms in the
potential well of its local cloud core. For a cluster mass of
$10^7\;M_\odot$ and a radius of 5 pc (G08), the virial speed in a
uniform sphere is $\sim40$ km s$^{-1}$. The velocity dispersion of the
gas can also be measured directly. The FWHM of the CO line around
cluster M4 is 70 km s$^{-1}$ (S07, Table 5), so the Gaussian dispersion
is 30 km s$^{-1}$, similar to the likely cluster dispersion. The CO gas
regions around M5 and M6 have similar dispersions (S07). The CO
velocity dispersion in the bar dustlane may be obtained from the
position-velocity diagram in Figure 6 of S07, and it is about the same,
30 km s$^{-1}$. We therefore assume the gas velocity dispersion is
$\sim30$ km s$^{-1}$.  This is high compared to observed dispersions in
the outer parts of disk galaxies. In the inner regions of NGC 1365
there is a deep potential well and a lot of energy available for gas
agitation in the motion of the bar and in the relative motions of gas
streams.

A velocity dispersion of $\sigma=30$ km s$^{-1}$ makes the cloud around
M4 close to virialized:  the virial mass inside the telescope beam with
radius $R=115$ pc is $M_V\sim5R\sigma^2/G= 1.2\times10^8\;M_\odot$,
which is a factor of only 2.2 higher than the CO-derived H$_2$ mass. We
can also set an upper limit to the CO cloud size using the extent of
the source in the narrow filter [NeII] 12.8\micro~image (G08). In this
image, the FWHM of the source is $\sim$0.5\arcsec. If we identify this
FWHM with the cloud diameter, then the radius would be 45 pc and the
virial mass $5.1\times10^{7}\;M_\odot$, which is the about same as the
CO mass.

The column densities can now be used to determine the gaseous scale
height. S07 estimated from the CO rotation curve that the stellar mass
inside the inner 1 kpc radius is $10^{10}\;M_\odot$, so the average
stellar mass column density is $3200\;M_\odot$ pc$^{-2}$. The molecular
gas layer is probably thinner than the stellar layer, so stellar
gravity adds to gas gravity in establishing the gas layer thickness.
For a two-fluid disk, the midplane gas pressure is approximately
(Elmegreen 1989)
\begin{equation}
P=\left(\pi/2\right)G\Sigma_{gas}\left(\Sigma_{gas}+{{\sigma_{gas}}\over{\sigma_{stars}}}
\Sigma_{stars}\right)\end{equation} and the midplane gas density is
$\rho_{gas,0}=P/\sigma_{gas}^2$. Here, $\Sigma$ is the mass column
density of either gas or stars, and $\sigma$ is the velocity
dispersion; $G$ is the gravitational constant. The scale height is
$H_{gas}=\Sigma_{gas}/\left(2\rho_{gas,0}\right)$, which is
\begin{equation}
H_{gas}\approx{{\sigma_{gas}^2}\over{\pi G \left(\Sigma_{gas}+
\left[{{\sigma_{gas}}/{\sigma_{stars}}}\right]
\Sigma_{stars}\right)}}.\end{equation} Setting
$\Sigma_{gas}=290\;M_\odot$ for the average value in the inner kpc,
$\Sigma_{stars}=3200\;M_\odot$ pc$^{-2}$, $\sigma_{gas}=30$ km
s$^{-1}$, and $\sigma_{stars}=100$ km s$^{-1}$ from Emsellem et al.
(2001), the gas scale height becomes $230/\left(1+3.3\right)\sim50$ pc.
It follows that the average midplane gas density is
$\rho_{gas,0}=2.9\;M_\odot$ pc$^{-3},$ which corresponds to an average
H$_2$ density of 50 cm$^{-3}$ in the inner kpc. The corresponding
pressure would be this density multiplied by $\sigma^2$, or
$1.3\times10^7k_B$ for Boltzmann's constant $k_B$. The stellar column
density dominates the gas in this evaluation of scale height, so in the
dense region surrounding the three clusters, where the average $H_2$
column density is $\sim500\;M_\odot$ pc$^{-2}$, the scale height is
about the same, $\sim45$ pc, although the average density is higher,
$5.5\;M_\odot$ pc$^{-3}$, or 95 H$_2$ cm$^{-3}$. The stellar scale
height is $H_{stars}=\sigma_{stars}^2/\left( \pi G\left[\Sigma_{gas}+
\Sigma_{stars}\right]\right)\sim210$ pc.

In summary, three $\sim10^7\;M_\odot$ clusters, $\sim7$ Myr old, are
each projected against molecular clouds in the ILR region of NGC 1365.
The clouds have masses of several$\times10^7\;M_\odot$ and column
densities of $500-900\;M_\odot$ pc$^{-2}$. They are part of a dense
massive dustlane in the northeast that connects the ILR region to the
outer part of the bar. A similar dustlane is in the south west part of
the bar. The total molecular mass in the bar is
$\sim3.8\times10^{9}\;M_\odot$, of which $\sim26$\% is in the inner kpc
region near the ILR. The velocity dispersion in the molecular gas is
$\sim30$ km s$^{-1}$, which suggests the individual clouds are
virialized, the average gas pressure in the inner kpc is $\sim10^7k_B$,
the gas scale height is $\sim50$ pc, and the average gas density is
$\sim50$ H$_2$ molecules cm$^{-3}$. In the CO plateau around the 3
clusters, the average H$_2$ density is $\sim95$ cm$^{-3}$.

\section{Where the clusters formed}
\label{sect:where}

The birthplace of the clusters can be assessed from the CO velocities,
the cluster ages, and the bar pattern speed. The clusters have
projected distances from the galactic center of 640 pc, 920 pc, and 760
pc (G08), respectively, which correspond to 800 pc to 900 pc in the
deprojected frame. According to Figure 6 in S07, the projected orbital
speed at this distance is 130 km s$^{-1}$ in the north-east and 160 km
s$^{-1}$ in the south-west, making the average speed 145 km s$^{-1}$.
We assume a value of 150 km s$^{-1}$ as in S07. For a 40\deg\
inclination, the deprojected orbital speed is 230 km s$^{-1}$. Then the
orbital time at 900 pc radius is 24 Myr.  The pattern speed, according
to L96, is 18 km s$^{-1}$ kpc$^{-1}$, so at the distance to the
clusters, the circular speed of the pattern is 16 km s$^{-1}$. Thus the
orbital speed relative to the pattern at the radius of the clusters is
$230-16=214$ km s$^{-1}$, and the orbital time relative to the pattern
is $2\pi\times900\;{\rm pc}/214\;{\rm km \;s}^{-1}=26$ Myr at 900 pc.

This relative orbital speed of 214 km s$^{-1}$ can be multiplied by the
7 Myr age of the clusters to get the total distance they would have
moved in circular orbits relative to the bar pattern. The result is
1500 pc, which corresponds to an angular displacement of 17\arcsec~at the
distance to NGC 1365, and to an arc around the center, for a circular
orbit, equal to 1.7 radians or 95\deg. This means that if the
clusters were moving in circular rotation around the center, then they
would have been born somewhere near the minor axis of the bar, which is
currently in the south-east -- they could not have formed in their
current dustlane. This possibility seems unacceptable because of their
clear proximity to the northern dustlane, which is one of the few
places where they could have formed, and because of the CO cloud at the
position of cluster M4. It is unlikely that this cloud stayed inside
the northern dustlane with the bar pattern speed and that the M4
cluster moved in a circular orbit with the orbital speed and just
happened to coincide with the cloud today. More likely, both the
clusters and the gas have not been moving in circular orbits.

We recently obtained SINFONI 2-D spectra for the three clusters. The
analysis of these data is in progress, and they will be fully presented
in a forthcoming paper (Galliano et al. in preparation). The spectra
show that the radial velocity in CO and \Bg~are equal: the CO clouds
are comoving with the clusters and they are all likely to be part of
the inward streaming dustlane flow.

The velocity field shown in Figure 12 of S07 indicates an inward
streaming speed along the bar dustlane of $\sim40$ km s$^{-1}$ in the
projected image at the position of M4. Similar streaming motions are
also observed in the velocity fields of optical emission lines
(Lindblad et al. 1996). To get the in-plane physical speed, this 40 km
s$^{-1}$ should be corrected upward by the factor
$1/\left(\sin\alpha\sin i\right)\sim2$, where $\alpha\sim50^\circ$ is
the pitch angle of the dustlane and $i\sim40^\circ$ is the galaxy
inclination. Thus the true speed of the inflow is more like 80 km
s$^{-1}$ along the dustlane.  This implies that in the last 7 Myr, the
gas in the dustlane near M4 traveled inward for $\sim560$ pc while it
rotated around at the pattern speed for $\sim110$ pc.  The inward
distance corresponds to an angular displacement of 6.2\arcsec~at the
distance of NGC 1365. In fact, the dustlane is still dense at a
position 6\arcsec~further out from M4, so this is the likely formation
site for M4 and the $5.4\times 10^7\;M_\odot$ cloud associated with it.
The same could be true for the other clusters, i.e., they all formed in
the north-eastern bar dustlane some 500-600 pc further out from their
current position at a galactocentric distance of $\sim1500$ pc
(17\arcsec from the center).

The suggested formation position for the clusters is highlighted in the
insert in Figure \ref{schematic_version2}. It is located inside the
formal ILR radius given by L96, which is indicated by an ellipse, and
outside the dust ring, which is at about half the ILR radius. The
overall flow of gas in this region is discussed next in order examine
possible origins for the clusters.

\section{Gas flow pattern in the bar and spirals}
\label{sect:flow}

\begin{figure}\epsscale{1}
\plotone{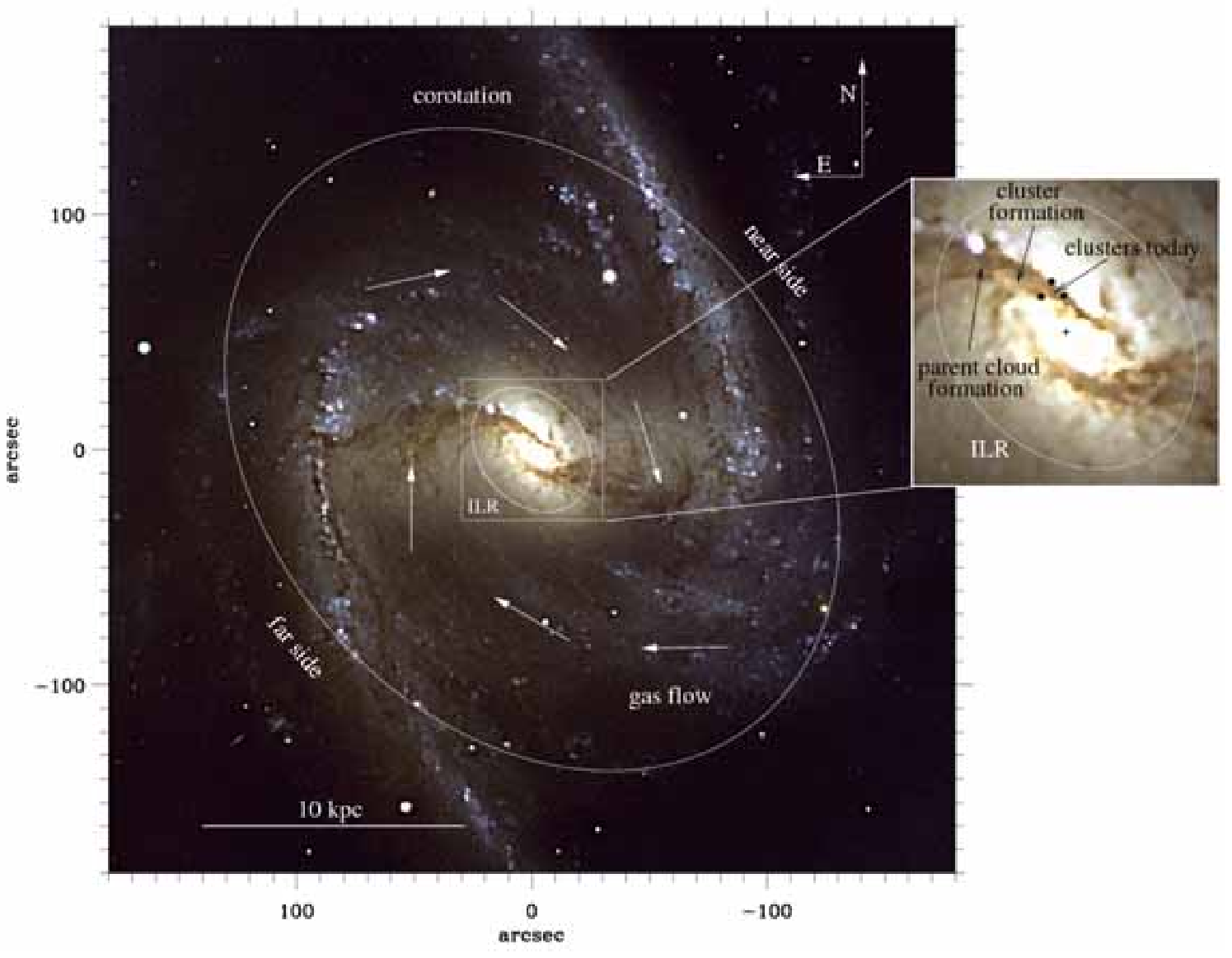} \caption{Optical image of NGC 1365 from three
exposures with FORS1: B(blue), V(green), and R(Red). Overlays show the
corotation radius, suggested flow lines based on dust filaments, the
current positions of the clusters M4, M5, and M6, and the suggested
formation positions of these clusters and their parent molecular cloud,
based on ages, dynamical times, and gas velocities. (Image degraded for
arXiv. see 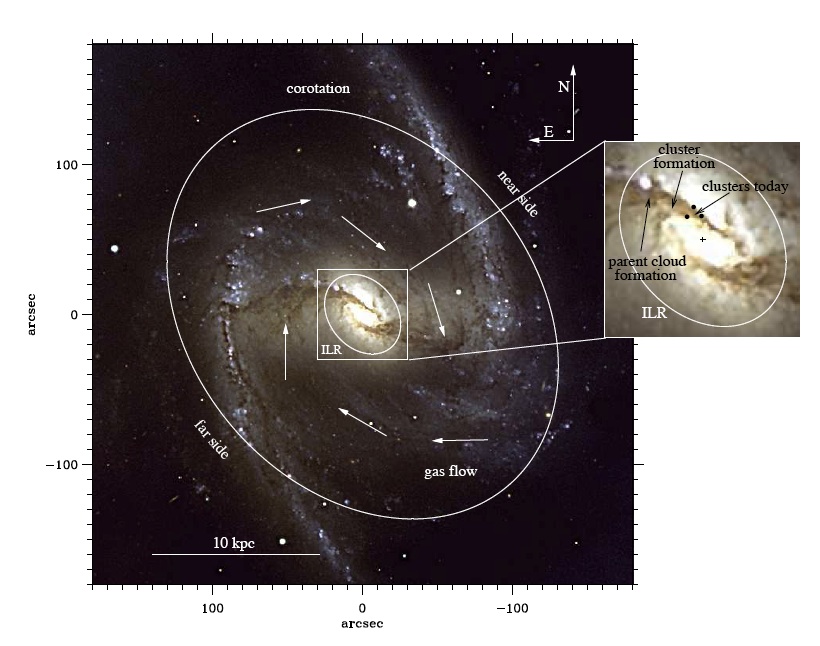)}\label{schematic_version2}\end{figure}

Figure \ref{schematic_version2} shows dense gas streams or filaments
intersecting the bar dustlane from the interbar region. Some of these
streams are traced by arrows. These dense streams have the form of arcs
that can be traced back to the spiral arms on the opposite side of the
galaxy. As corotation for the bar seems to be 1.31 bar lengths (L96),
the brightest spiral region just outside the bar is still inside
corotation. Most of the gas there should be moving inward as a result
of bar and spiral torques.

The dust structure in Figure \ref{schematic_version2} suggests that the
spiral arm gas does not deflect sharply at the bar end and move inward
along the bar dustlane. The spiral arm dust moves from the spirals into
the leading interbar region along filaments, and then it impacts the
bar dustlane after curving back around at a smaller radius. Perhaps an
impact like this between a small stream and the bar dustlane triggered
the formation of clusters M4, M5, and M6. Each dust stream in the
interbar region can be traced back through filaments to the spiral arm
on the other side of the galaxy.  The suggested formation site of the
parent cloud for the clusters, discussed in more detail in section
\ref{sect:formation}, is indicated in Figure \ref{schematic_version2}.
It is indeed at a place in the bar dustlane that is intersected by a
filament from the interbar region.

In the south-east, the spiral arm bifurcates at about the corotation
radius (Figure \ref{schematic_version2}). The outer part of this
bifurcation is beyond corotation for the pattern speed given by L96, so
the gas and stars there should eventually move outward because of bar
and spiral torques (Lynden-Bell \& Kalnajs 1972; Schwarz 1981). The
inner part of this bifurcation, along with the brightest parts of the
spiral arms, are inside corotation and they should eventually move
inward to joint the bar dustlane and ILR ring. Considering the
in-streaming speed of $\sim80$ km s$^{-1}$ along the bar dustlane (see
above), and the deprojected bar radius of 120\arcsec~or 10.8 kpc for
our assumed distance (JM95, L96), the time for the gas in the bar
dustlane to fall in from the spiral position to the ILR is $\sim130$
Myr, which is 2.4 pattern rotation times (using $\Omega_p=18$ km
s$^{-1}$ kpc$^{-1}$ from L96).

The gas flow model proposed here and based on the distribution of dust
filaments in Figure \ref{schematic_version2} differs from the
simulations in L96 in two important ways. First, their modeled gas flow
in the bar region (their Figure 26b) has gas in one bar dustlane flow
down along the dustlane and then out in the prograde direction to the
other dustlane, where it shocks and gets compressed again. Then it
flows out of the other dustlane in the prograde direction and back to
the first dustlane. In this way, the gas circulates to the center
slowly, taking 3 or more rotations relative to the bar to decrease its
radius by a factor of 2 (in their Figure 26b; see also Athanassoula
1992). As a result, the net accretion rate of the dustlane gas can be
small. Alternatively, the gas in the bar dustlane can go directly to
the inner kpc region without emerging on the prograde side of the bar
and hitting the other dustlane (e.g., Piner, Stone \& Teuben 1995;
Regan, Sheth, \& Vogel 1999).

The primary reason for forward gas emergence out of the dustlane in
some bar flow models is the assumption of an isothermal or other
simplified gas equation of state. When gas is isothermal, for example,
the pressure in the dustlane is large, proportional to the density, and
the leading edge of the bar dustlane has a pressure gradient that
accelerates the gas ahead of the lane. However, the assumption of an
isothermal or other simplified gas is inappropriate for this situation.
An isothermal gas has the property that every decompression has a
source of energy to keep the temperature constant, but in fact there is
no such source for a bar dustlane. The gas enters the dustlane, cools
rapidly by CO and other molecular emission, and stays cold. There
should be no significant pressure gradient at the front edge, and there
should be no source of energy that would allow this gas to accelerate
significantly away from the dustlane in the forward direction. The gas
should fall to the central region along the bar dustlane and speed up
as it moves in. Note that once the gas is in the dustlane and streaming
inward, its angular momentum is already as low as it has to be to reach
the inner dust ring by direct collapse.  Further accretion to the
nucleus depends on details of the bar potential and whether there is a
nuclear bar (e.g., Knapen et al. 2002; Regan \& Teuben 2004).

The assumption of an isothermal gas is useful for studies dominated by
gas compression, because energy loss during compression can keep the
temperature or velocity dispersion in the gas about constant. However,
this assumption is inappropriate for decompression, because there is no
inverse process of energy absorption to keep the velocity dispersion
constant during the decompression. Interstellar shocks are a one-way
path toward high density, especially if the magnetic field diffuses out
rapidly in the compressed phase. The dense regions can only be broken
into pieces by other shocks or clouds and only slowly evaporated to a
lower density by ionization and thermal heating.

The second difference between our model and the simulations in L96 is
that the spiral arm gas in L96 moves from one spiral arm to the next,
losing only a small amount of orbital energy with each cycle. In Figure
\ref{schematic_version2}, however, one sees dense dust filaments almost
radially aligned that connect the spiral arms with the bar dustlane,
suggesting a gas flow directly from the spiral region to the bar, in
just one-half of an orbit (in the pattern frame). A few examples of
this type of flow are in Regan \& Teuben (2004) for their thick bar
cases (their Fig. 5).

\begin{figure}\epsscale{1}
\plotone{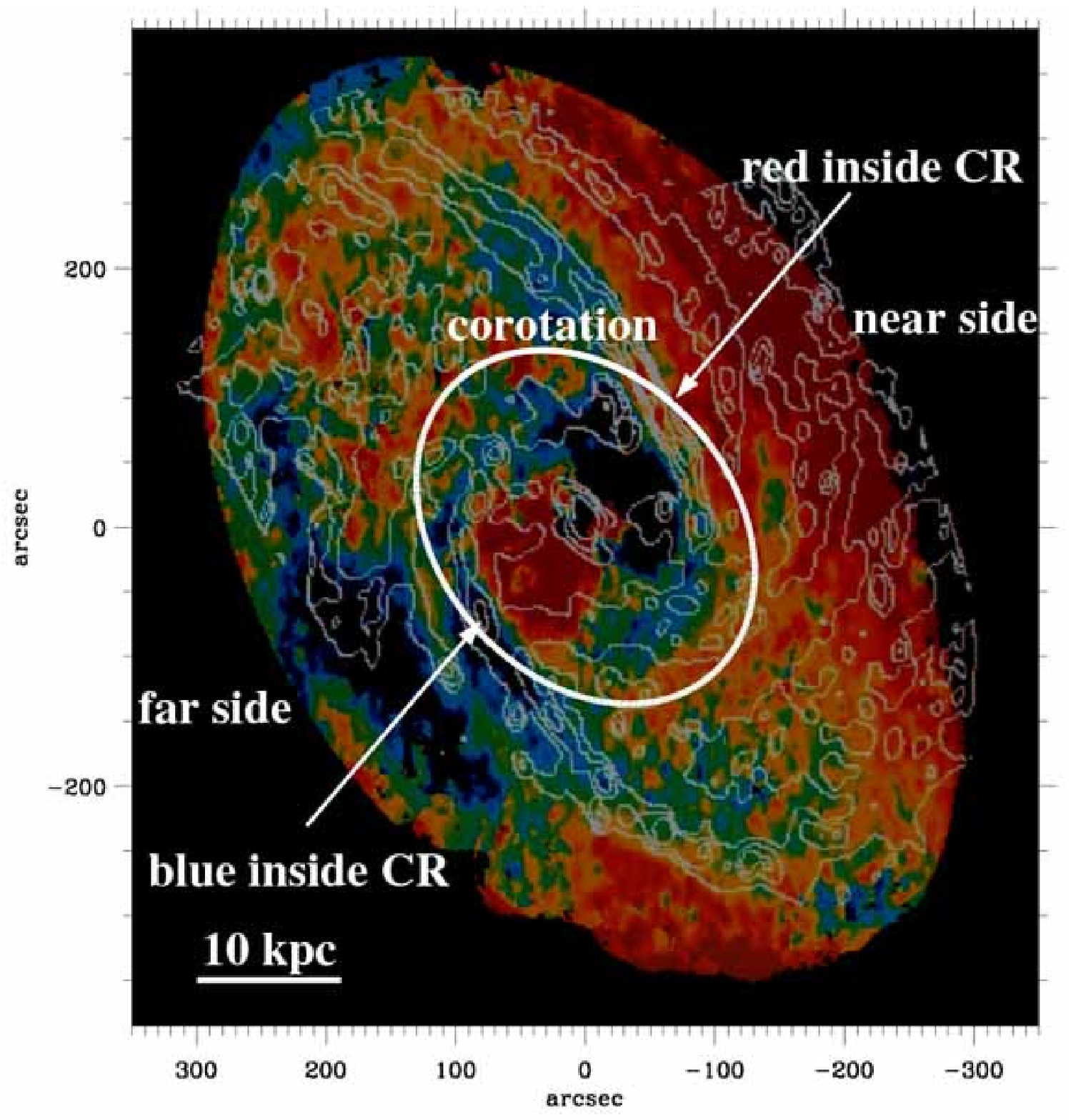} \caption{Reproduction of the HI velocities (color)
and B-band optical intensities (contours) of NGC 1365 from Figure 17 of
JM95, with notation added. The color scale ranges from $-25$ km
s$^{-1}$ in the extreme blue to $+25$ km s$^{-1}$ in the extreme red.
The sign of spiral arm streaming changes at corotation from inward
inside corotation to outward outside corotation, as indicated by the
color change. (Image degraded for arXiv. see
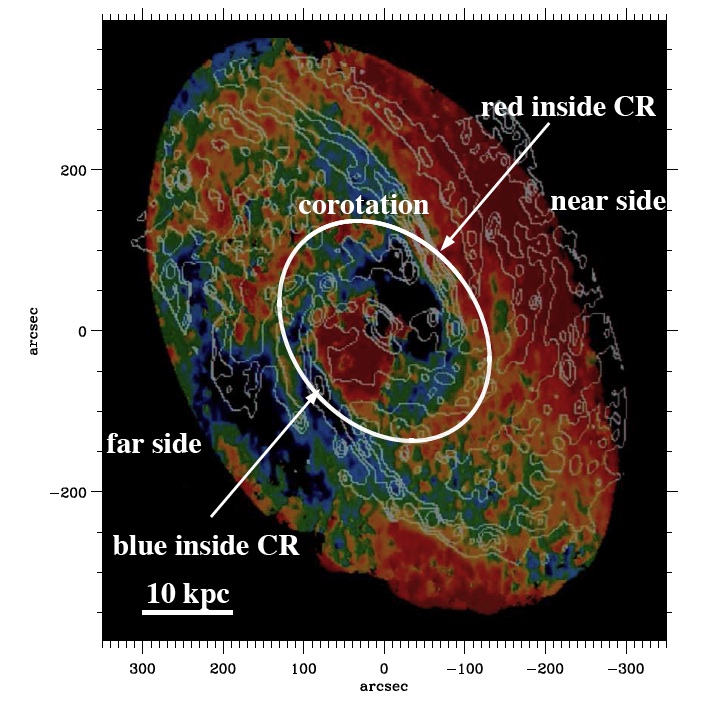.)}\label{fig-Jorsater95}\end{figure}

Spiral arm streaming motions can be seen in the HI velocity map
presented by JM95 (their Figures 9 and 17). Their Fig. 17 is reproduced
in color here in Figure \ref{fig-Jorsater95} because the on-line
version of that paper (NASA Astrophysics Data System Bibliographic
Services) has a Black and White image. On the northwest minor axis,
there is a transition from positive line-of-sight velocities (red)
inside corotation to negative velocities (blue) outside corotation in
the same spiral arm.  Because this is the near side of the galaxy, the
positive velocities inside corotation are inward. On the southeast
minor axis, the spiral arm velocities just inside corotation are
negative (blue), which is also inward on this far side position. JM95
also note a spiral arm inflow near the bar end. From Figures 17 and 21
in JM95 (Fig. \ref{fig-Jorsater95} here), the line-of-sight inflow
speed is determined to be about $-15$ km s$^{-1}$ on the near and
far-side minor axes. Because these are the minor axes, the component of
the inflow speed in the direction of the galactic center is $15/\sin
i=23$ km s$^{-1}$ for an inclination of $i=40^\circ$.  The full flow
speed along the spiral arm is this 23 km s$^{-1}$ radial component
divided by the sine of the deprojected pitch angle of the arm, which we
estimate to be $\sim25^\circ$ from the deprojected image of NGC 1365 in
JM95.  The full, parallel-to-arm streaming speed is thus $\sim50$ km
s$^{-1}$.

In summary, the gas in NGC 1365 is observed to stream outward outside
of corotation and inward inside of corotation, as expected from
numerous models and observations of other galaxies. In the spiral
region, the gas is mostly HI and it streams along the arms inside of
corotation at a radial speed of $\sim23$ km s$^{-1}$. Some of this
inflow apparently feeds dust filaments that impact and add to the bar
dust lane after circling halfway around the bar pattern. The bar inflow
speed is $\sim80$ km s$^{-1}$. Gas inflow should be faster and more
direct than simulations suggest if the equation of state for gas does
not artificially inject energy at rarefaction fronts.

\section{Gas Accretion Rates, Star Formation Rates, and the Age of the
Bar}\label{sect:accretion}

There is a large reservoir of atomic gas outside the bar in NGC 1365,
mostly in the spiral arms. According to Figure 4 in JM95, the projected
HI column density in the northern spiral arm between the bar-end and
corotation averages $\sim13\;M_\odot$ pc$^{-2}$. The width of the arm
is $\sim20$\arcsec~or 1.8 kpc, and the length is $\sim200$\arcsec, or
18 kpc, so the projected area is $3.3\times10^7$ pc$^{2}$. The HI mass
in this part of the spiral is therefore $\sim4\times10^8\;M_\odot$
(independent of projection effects). The inner part of the spiral at
the south-western end of the bar has a similar HI mass. Thus the total
HI mass available for accretion inside corotation is
$\sim8\times10^8\;M_\odot$ from the spiral arms alone and more from the
inter-arm regions.

The radial distance the spiral arm gas has to go before reaching the
bar is the spiral arm length between corotation and the bar, or
$\sim3.3$ kpc, considering a bar radius of 10.8 kpc and a corotation
radius of 1.31 bar lengths (L96). If the radial inward speed is
$\sim23$ km s$^{-1}$ along the arms, as determined in the previous
section, then the inflow time to the bar is the ratio of the radial
distance to the radial speed, or 140 Myr. The accretion rate into the
bar region from the spirals is the ratio of the accreting HI mass to
the accretion time, or $\sim5.7\;M_\odot$ yr$^{-1}$. There could be
more accretion if the gas between the spiral arms moved inward, or less
if it moved outward, but these flows are unobserved.

The accretion rate along the bar dustlane can be determined in a
similar way using CO data in S07. The molecular mass column density in
the north-eastern dustlane increases from $\sim40\;M_\odot$ pc$^{-2}$
at large radius to $\sim600\;M_\odot$ pc$^{-2}$ in the vicinity of the
clusters M4, M5, and M6. These are projected column densities, and they
should be decreased by the factor $\cos40^\circ$ to get perpendicular
surface densities. The increase in column density with decreasing
distance along the bar dustlane is consistent with the addition of gas
along the dustlane from the interbar filaments, as discussed in section
\ref{sect:flow}. The outermost part of the bar dustlane gets its gas
partly from the local spiral and interbar region, and partly from the
corotation region of the spiral arm on the other side of the galaxy,
while the inner part of the bar dustlane gets its gas from the inner
region of the other-side spiral arm plus the inflow along the dust
lane.

The radial inflow speed at small radius, near the three clusters, is
$\sim80$ km s$^{-1}$, as discussed in section \ref{sect:where}. The
projected width of the bar dustlane near the clusters is 5\arcsec~(see
Figure 2 in G08), which is 450 pc in the direction of the minor axis.
The inflow rate near the clusters is approximately the product of the
projected mass column density, the projected width, and the radial
inflow speed, which gives $\sim22\;M_\odot$ yr$^{-1}$ for the
north-eastern dustlane. For the same width and accretion speed in the
outer part of this dustlane, the accretion rate there is only
$\sim1.5\;M_\odot$ yr$^{-1}$ -- lower because of the lower column
density. More likely, the accretion rate in the outer part of the bar
dustlane is even lower than this because the gas inflow speed is lower:
the gas has not yet accelerated much into the potential well of the
galactic center. Midway in the bar dustlane, the molecular mass column
density is $\sim200\;M_\odot$ pc$^{-2}$, while the width and speed are
still around 450 pc and 80 km s$^{-1}$, so the accretion rate is
$\sim7\;M_\odot$ yr$^{-1}$. All of these rates should be doubled to get
the total accretion rate to the center, considering the similar
dustlane on the south-western side of the bar.

There is evidently a disconnection between the rate at which the spiral
arms can add mass to the outer bar region ($\sim5.7\;M_\odot$
yr$^{-1}$) and the rate at which the inner part of the bar dustlane
adds mass to the inner kpc ($\sim44\;M_\odot$ yr$^{-1}$). These rates
differ by a factor of $\sim8$. The spiral and bar accretion rates agree
with each other fairly well in the outer part of the bar (5.7 and
$3.0\;M_\odot$ yr$^{-1}$, respectively), but not in the inner part. The
accretion either increases by the addition of mass on the way in (e.g.,
in the observed interbar dust filaments), or the flow is not in a
steady state.

Using the star formation rate to far-IR luminosity relation of Kennicut
(1998) and considering that half of the far-IR luminosity of the galaxy
is due to the starburst, the star formation rate in the region of the
ILR was estimated to be $\sim9\;M_\odot$ yr$^{-1}$ (S07). This converts
to $\sim9.7\;M_\odot$ yr$^{-1}$ for our slightly larger distance. For
the total central gas mass of $9.7\times10^8\;M_\odot$, the gas
consumption time is short, 100 Myr (S07). However, the current nuclear
accretion rate is $\sim44\;M_\odot$ yr$^{-1}$, which is about four
times the star formation rate. Star formation occurs along the
dustlane, as we have seen for M4, M5, and M6, converting some of the
gas into stars before it reaches the center. Over the next several
hundred Myr, the gas accretion rate in the center will have to decrease
to the feeding rate by the spiral arms, which is $\sim5.7\;M_\odot$
yr$^{-1}$.

Eventually all of the gas inside corotation will reach the center and
either turn into stars or get expelled from the region by starburst
pressures. This can take longer than the current gas consumption time
if the bar slows down and grows in length over time. Such bar evolution
is expected because of dynamical friction on the bar exerted by the
disk and halo (e.g., Athanassoula 2003, and references therein). The
average HI surface density outside the bar between 100\arcsec~and
300\arcsec~radius is $\Sigma\sim7\times10^{20}$ H
cm$^{-2}\sim6.7\;M_\odot$ pc$^{-2}$, according to Figure 8 in JM95. If
the bar grows at a speed $v$, in km s$^{-1}$, then this gas is added to
the bar region at the rate $2\pi R\Sigma v\sim0.4v\;M_\odot$ yr$^{-1}$
for a radius comparable to the bar radius of $R\sim10$ kpc. In
Athanassoula (2003), bars can slow their pattern speed by a factor of 2
or 3 in $10^{10}$ years, which means their corotation radii increase by
this factor in the same time. To maintain a steady accretion rate at
the end of the bar that is equal to what we derived above,
$5.7\;M_\odot$ yr$^{-1}$, the corotation radius would have to grow at
an average rate of 14 km s$^{-1}$. This means it would have to increase
by 50\%, from 10 kpc to 15 kpc, in the next 0.3 Gyr. This is faster
than the corotation growth rate in the models by Athanassoula (2003),
which is more like 1 km s$^{-1}$ for a big bar.

Interactions with other galaxies could also maintain the current
accretion rate to the bar region. An interaction can drive in gas from
the far outer disk along tidal arms, and it can cause the bar to grow
more rapidly. Perhaps a grazing collision $\sim1$ Gyr ago led to the
present epoch of accretion and star formation in the center.

In summary, the nuclear region of NGC 1365 is currently accreting
matter at a total rate of $\sim44\;M_\odot$ yr$^{-1}$ along two dense
dustlanes. This inflow adds mass to the $9.7\times10^8\;M_\odot$ of
molecules that is already in the ILR region, and is enough to sustain
the star formation rate of $\sim9.7\;M_\odot$ yr$^{-1}$ for
considerably longer than the current ILR gas consumption time of
$\sim100$ Myr. If we consider that the total CO mass in the bar region
is $3.8\times10^9\;M_\odot$, from section \ref{sect:env}, and the inner
kpc already contains $0.97\times10^9\;M_\odot$, then an additional gas
mass of $2.8\times10^9\;M_\odot$ has yet to accrete from the bar
dustlanes to the central kpc. At an accretion rate of $\sim44\;M_\odot$
yr$^{-1}$, this additional gas would take an additional $\sim63$ Myr to
reach the ILR region. At the current star formation rate, the
consumption time for this total molecular gas would be $\sim390$ Myr.

The central inflow rate of $\sim44\;M_\odot$ yr$^{-1}$ has no source
this large in the outer part of the bar. The source there is primarily
the HI that is inside corotation, and the inflow rate there is only
$\sim5.7\;M_\odot$ yr$^{-1}$. At this low inflow rate, the HI reservoir
inside corotation of $\sim8\times10^8\;M_\odot$ will last 140 Myr.
After this 140 Myr, the HI mass inside corotation will have been added
to the current $3.8\times10^9\;M_\odot$ of molecules in the bar region,
giving a total gas mass for star formation inside corotation equal to
$4.6\times10^9\;M_\odot$. With an ILR flow rate of $\sim44\;M_\odot$
yr$^{-1}$, this total comes into the center in 105 Myr. At the current
star formation rate of $\sim9.7\;M_\odot$ yr$^{-1}$, an amount of gas
equal to $1.4\times10^9\;M_\odot$ will be converted into stars in the
inflow time of the HI reservoir near corotation, 140 Myr, and there
will still be $3.2\times10^9\;M_\odot$ of gas left over in the bar and
central regions. This gas can continue to form stars at the same rate
for another 330 Myr. Thus the current starburst can last for
$140+330=470$ Myr (or $4.6\times10^9\;M_\odot/9.7\;M_\odot\;{\rm
yr}^{-1}=470$ Myr), building up an additional stellar mass in the ILR
and bulge region of $4.6\times10^9\;M_\odot$, which is the total gas
mass currently inside corotation. After that, the star formation rate
can only be comparable to the accretion rate in the outer part of the
bar from the gradual growth of the bar length. Bar growth at a likely
$\sim1$ km s$^{-1}$ will add gas mass to the corotation region and
ultimately to the ILR region at a rate of $0.4\;M_\odot$ yr$^{-1}$.
This will likely be the star formation rate in the ILR region after the
470 Myr period of rapid accretion is over.

Most likely, the bar is not much older than this gas consumption time,
perhaps 1-2 Gyr, because there is no long-term source of gas from
outside the bar region that could have maintained the current gas
supply much longer than this. The gas was presumably inside corotation
when the bar formed and it has been accreting and forming stars in the
ILR region ever since. Because the current accretion rate exceeds the
current star formation rate in the center, the bar in NGC 1365 could
still be in a youthful phase where it is cleaning out the gas that was
formerly in a bar-less inner galaxy disk. The central starburst could
get even stronger in the next hundred million years as the rest of the
bar gas comes in.

A similar time for star formation, $\sim0.5$ Gyr, was suggested for the
ILR ring region of the galaxy M100 by Allard et al. (2006). The bar
there could be young too (1-2 Gyr) for the same reasons as given here,
i.e., a lack of gas from other reservoirs.  Bar ages and star formation
timescales might be longer if gas from the main disk outside of
corotation can also get into the bar region.  Sorai et al. (2000)
suggested that viscous forces might do this. Bars without active
nuclear star formation would have no such constraints on their ages.

Gas accretion and star formation rates have been estimated for several
other barred galaxies. Normally they are smaller than the values for
NGC 1365.  Meier, Turner, \& Hurt (2008) measured an accretion rate of
$\sim0.7\;M_\odot$ yr$^{-1}$ in Maffei 2, and noted that this was
higher than the ILR star formation rate by a factor of 5. This result
is similar to ours, suggesting sustained star formation and further gas
buildup, but it is scaled down in Maffei 2 by a factor of 10 to 100
because of the smaller ILR radius ($\sim100$ pc compared to $\sim1$
kpc).  Wong \& Blitz (2000) obtained inflow speeds for NGC 4736 of
several tens of km s$^{-1}$ and an accretion rate of $\sim2\;M_\odot$
yr$^{-1}$, which is $\sim10\times$ higher than the inner ring star
formation rate in that galaxy. Martin \& Friedli (1997) derived an
accretion rate $\sim4$ times higher than the star formation rate in NGC
7479, which was one of the first galaxies to have an estimated
accretion rate, given by Quillen et al. (1995) as $\sim4\;M_\odot$
yr$^{-1}$. Regan et al. (1997) obtained an accretion rate for NGC 1530
of $\sim1\;M_\odot$ yr$^{-1}$.  Evidently, the barred galaxy studied
here, NGC 1365, is unusually large and gas-rich, having a very high
column density of molecules in the inner star-forming region and a high
total gas mass inside corotation. Still, it is difficult to see how it
could have sustained its current level of activity for a Hubble time,
given the available gas and the current rate of star formation.

We conclude that gas accretion and star formation in the ILR region of
NGC 1365 is most likely variable by a factor of $\sim3$ or more, as
gated by the release of gas near corotation for accretion into the
center, and that the total duration of the current burst might be only
$\sim0.5$ Gyr. The bar itself is probably not much older. Bar accretion
is apparently along the bar dustlane and also along filamentary flows
that leave the spiral arm region inside corotation and loop into the
bar dustlane where they get assimilated. The dustlane presumably
accumulates matter in this way all along its straight path to the ILR.
Star formation may be triggered at the intersection points between the
dustlane and the looping filaments, and it may also be triggered by
spontaneous gravitational instabilities in the dustlane (section
\ref{sect:formation}). Because of high pressures in the intersection
points, dense massive clusters can form.

\section{On the Cluster Mass Function}\label{sect:cmf}

One peculiarity of the clusters is their large mass,
$\sim10^7\;M_\odot$. Usually when clusters of this mass are found, the
total mass in all clusters formed at about the same time is $\sim15$
times larger for a typical cluster mass function $dN/dM\propto M^{-2}$.
That is, the total mass of clusters between a minimum mass $M_n$ and a
maximum mass $M_x\sim10^7\;M_\odot$ is $M_x\ln(M_x/M_n)\sim15 M_x$ for
$M_n\sim10\;M_\odot$. This implies that the total mass of clusters with
comparable age in the same region should be
$\sim1.5\times10^8\;M_\odot$, or perhaps three times larger considering
there are three visible clusters each with a mass around
$10^7\;M_\odot$. Recall that the total gas mass inside the entire 1 kpc
radius of the galaxy is $9.7\times10^8\;M_\odot$, so the total cluster
mass, considering the factor of 3, would be half of the total current
gas mass. The gas mass in the region immediately surrounding the three
massive clusters is $\sim10^8\;M_\odot$ from the CO(1-0) contours in
Figure 2 of S07 (Sect. \ref{sect:env} here). This is such a small gas
mass that essentially all of the other clusters that should be there in
a normal cluster mass function would have had to form with nearly 100\%
efficiency. Moreover, these clusters would have to be obscured by dust
because only the three $10^7\;M_\odot$ clusters are prominent. Such
obscuration is not inconceivable because the visual extinction
corresponding to the average H$_2$ column density around the clusters,
$\sim500\;M_\odot$ pc$^{-2}$ (S07) is $\sim35$ mag -- large enough to
hide lower luminosity clusters. Thus we attempt to determine if there
is a complement of clusters at, for example, one-tenth the mass of the
observed three clusters, in the same dense dust lane with about the
same age.  According to the cluster mass function, there should be ten
times as many clusters with one-tenth the mass of M4, M5, and M6.

\begin{figure}\epsscale{1}
\plotone{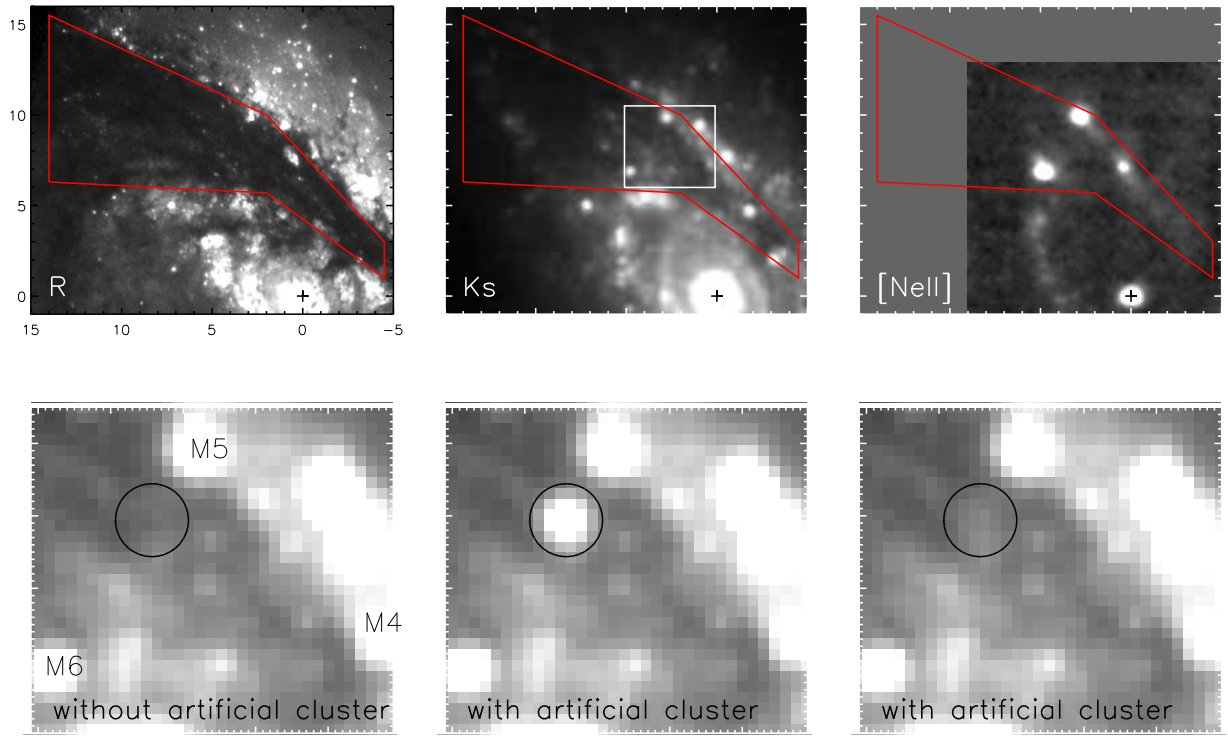} \caption{The top three panels show the inner region of
NGC 1365 surrounding the bar dust lane in the north in R, Ks, and
[NeII] 12.8\micro~narrow filter images. The active galactic nucleus is
identified with a cross. The bottom three panels show an enlargement in
the Ks band of the region inside the white rectangle in the top where
the three massive clusters are. The bottom left shows the raw image,
the bottom center shows the same image but with the addition of an
artificial source having the same flux as M6.  The bottom right has an
artificial source with 1/10th the flux of M6.  The 1/10th artificial
source is barely detectable, as are similarly faint real sources in
this image. Evidently, there are several tens of clusters in this
region with masses of $\sim0.1$ times the mass of M6 if they all have
the same age.} \label{clusters}\end{figure}

We searched for these 30 expected clusters in infrared images.  The
region is shown in Figure \ref{clusters} at three different
wavelengths: on the top, from left to right are displayed the R, Ks,
and [NeII] 12.8\micro~narrow filter images (from G08). The dustlane
region where we should search such a population of fainter clusters is
indicated by a red boundary. In the visible image (R), only a few
sources are detected in this region.  Because the extinction can be
very large at this wavelength, this image is not well suited for
looking for the fainter clusters.

In the K-band, the extinction is much lower. Inside the region of
interest we can see faint sources between the three massive clusters.
Measuring the fluxes of these sources is challenging because of
crowding. In order to estimate how detectable a cluster is in this
image, we have added an artificial source of known flux. On the bottom
row of Figure~\ref{clusters} are shown three versions of the zoomed M4,
M5, M6 region of the Ks image. This zoomed region is delineated with a
white rectangle on the top middle (Ks) image. On the bottom row: to the
left, the ``raw'' image is displayed without any added sources, in the
middle, an artificial source with the same flux as M6 has been added
(within the black circle), and to the right, an artificial source with
1/10th the flux of M6 has been added (within the black circle). We see
that, in the latter case, the added source is at the limit of
detection. However, in the original K-band image, there appear to be
several tens of similar faint sources in the area of the dense
dustlane. Thus, these faint sources may represent the $\sim
10^6\;M_\odot$ ~part of the same cluster mass function that formed M4,
M5, and M6.

If these fainter clusters had the same emission line intensities as M4,
M5, or M6, then we should detect their \Pa~and \Bg~line emission along
the long slit 2\micro~spectra presented in G08. In fact, we do not
detect this line emission.  The lack of detection suggests that the
fainter clusters have spectra with proportionally fainter nebular line
emission. This observation is also consistent with the narrow band
[NeII] 12.8\micro~image (Figure \ref{clusters}, top row, right map), in
which no population of faint sources can be seen. On this map, the
fluxes of the three massive clusters are in the range 100 to 300\,mJy,
and the detection limit is down to 10\,mJy. This means that there are
no other [NeII] emitting clusters in this region down to at least
$10^6\;M_\odot$, if the line emission is proportional to the continuum.

In summary, the dustlane region contains several tens of sources that
have K-band fluxes of about 1/10th that of the most massive clusters.
If these clusters formed at the same time as the massive clusters, then
the initial cluster mass function could be normal, The fainter clusters
do not exhibit emission line fluxes quite in proportion to their
infrared luminosities, however. If they are young, then their lack of
[NeII] lines suggests that they removed most of their residual gas,
unlike the more massive clusters which still have prominent [NeII]
emission.  We consider possible reasons for this difference in the next
section.

Alternatively, the fainter clusters could be older than the massive
clusters, perhaps from earlier star formation in the ILR region. Old
clusters have been found alongside young clusters in the ILR region of
M100 (Allard et al. 2006).  If they are indeed older, then these
clusters must also be extremely massive because of their fading with
age.  For ages between 15 and 40 Myr, the faint clusters would be about
as massive as the three younger ones (from Leitherer et al. 1999). Then
the northern dust lane would contain $\sim30\times10^7\;M_\odot$ of
clusters. In 40 Myr, the star formation rate required is $7.5\;M_\odot$
yr$^{-1}$, which is close to the observed rate of $\sim9.7\;M_\odot$
(S07). In this case, there would be no obvious population of low mass
clusters to make a power-law initial cluster mass function.

\section{Gas Removal from Clusters}\label{sect:remove}

The presence of [NeII] line emission from the most massive clusters in
the inner region of NGC 1365 and the lack of any obvious emission from
fainter clusters nearby could be the result of detection limitations
for the fainter clusters.  If, however, the low-mass clusters really do
have significantly less ionized gas than the massive clusters, even
less than the expected proportion to their luminosity, then this
observation is somewhat peculiar. Usually, massive clusters are
expected to have higher pressures for clearing away their natal gas,
and so less ionized gas in proportion to their mass than lower mass
clusters. Clearing depends also on gravitational self-binding of the
gas, however, and it could be that lower-mass clusters have more weakly
bound gas for their pressures than massive clusters.

Here we compare the energy released by a cluster's stellar winds and
supernovae to the binding energy of the residual star-forming gas. The
energy released by the stars increases proportionally to the cluster
mass, while the gravitational binding energy of the gas in the cluster
increases like the square of the cluster mass for a fixed efficiency.
Therefore, low-mass clusters should be able to clear away their gas
more effectively than high-mass clusters. At some critical minimum
mass, a cluster should have great difficulty removing its gas (Murray
2009). A similar mass dependence for gas expulsion that causes cluster
disruption was considered by Baumgardt, Kroupa \& Parmentier (2008).
The tendency for high mass clusters to accrete gas from their
environments was discussed by Pflamm-Altenburg \& Kroupa (2009).

For the energy released by a star cluster, we consider stellar winds
and supernovae with an efficiency for gas removing of 10\% (MacLow \&
McCray 1988). For the ability of gas clearing by ionizing radiation, we
consider a smaller efficiency of 0.1\% (Dale et al. 2005). The time
dependencies of the cluster energy outputs are computed from
Starburst99 (Leitherer et al. 1999).

The energy necessary to expel gas from a cluster is approximately the
binding energy,
\begin{equation}
E_{\rm grav}={{ G M_{\rm gas}^2 }\over{2R}} + {{GM_{\rm star}M_{\rm
gas}}\over R }.
\end{equation}
The factor of 2 in the denominator of the first term in the equation
comes from the reduction in the potential as the gas particles are
removed.

\begin{figure}\epsscale{1}
\plotone{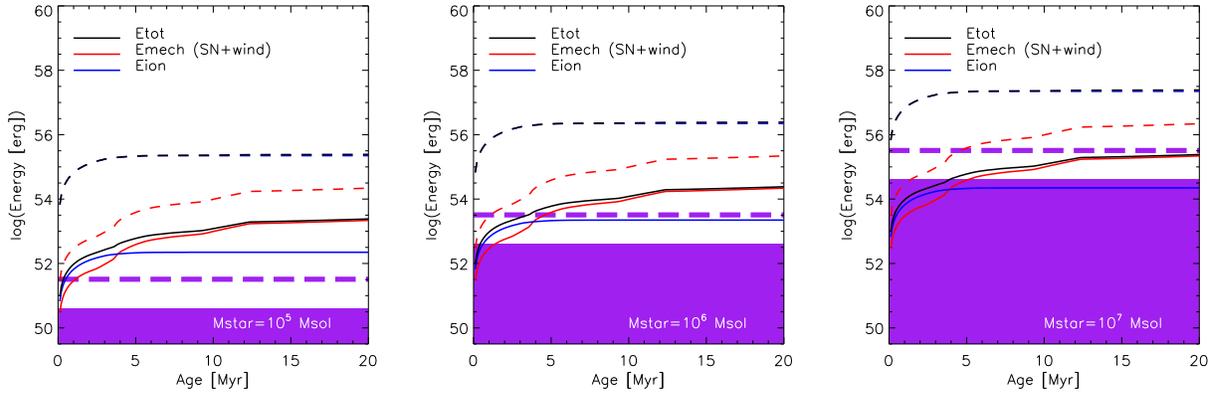} \caption{Models showing the energy sources and sinks
for clusters with masses of $10^5$, $10^6$, and $10^7\;M_\odot$, from
left to right. The dashed lines are the total energies put out by the
clusters in the three forms indicated: total, mechanical (supernovae
and stellar winds), and ionization. The solid lines are these energies
multiplied by the efficiencies for pushing cluster gas away. The
gravitational binding energy is shown in purple, with shading for
R=40\,pc and a thick dashed line for R=5\,pc. The gas mass is assumed
to be 2.3 times the star mass. This figure suggests that low mass
clusters put out an amount of energy that can clear the gas away within
only a million years, while a $10^7\;M_\odot$ cluster cannot clear the
gas away for $\sim7$ Myr. This result may explain why the three massive
clusters still emit [NeII] while the low mass clusters nearby do
not.}\label{cluster_energy}\end{figure}

Figure \ref{cluster_energy} plots the time dependence of the total
cluster energy released multiplied by the efficiencies given above (up
to 20 Myr). The gravitational binding energy, $E_{\rm grav}$, is
indicated by the shaded region. Three cluster masses are considered
from left to right, $M_{\rm star}=10^5$, $10^6$, and $10^7\;M_\odot$.
The star formation efficiency is taken to be 30\%. For the gas
distribution radius, we take two values: (1) R=5\,pc, in which case we
consider that the gas is contained inside the radius of the stellar
cluster, as derived from the HST image (G08); (2) R=40\,pc, where we
consider that the gas bound to the cluster can have a more extended
distribution. The gravitational binding energy, $E_{\rm grav}$, is
indicated in purple using shading for R=40\,pc and a thick dashed
purple line for R=5\,pc. The dashed lines show the total energy values
and the solid lines are the energies multiplied by the efficiencies
(0.1 for mechanical energy and 0.001 for ionization energy). The total
cluster energy released is the black dashed line.

Figure \ref{cluster_energy} suggests that at a cluster mass of
$\sim10^7\;M_\odot$, the binding energy is larger than the total
clearing energy available from the cluster up to $\sim7$ Myr, whereas
the lower mass clusters are cleared of gas more quickly. The critical
mass mentioned earlier would then be $\sim10^7\;M_\odot$. This could
explain why the three most massive clusters have not expelled their
surrounding gas yet, causing them to exhibit intense [NeII] line
emission. In contrast, the $\sim10^6\;M_\odot$ mass clusters  in NGC
1365 might have been able to clear their surrounding gas in their short
lifetimes, and for this reason do not show detectable [NeII] line
emission.

\section{Cluster Formation}\label{sect:formation}

The characteristic mass and separation of the largest regions of star
formation in a gas disk should be comparable to the Jeans mass and
Jeans length. These scales come from the dispersion relation for
gravitational instabilities in a thin galaxy disk,
$\omega^2=k^2\sigma^2-2\pi G\Sigma k+\omega_{ep}^2$ for rate $\omega$,
wavenumber $k$ and epicyclic frequency $\omega_{ep}$. The wavenumber of
fastest growth is $k_J=\pi G\Sigma/\sigma^2$, half the wavelength is
$\lambda_J/2=\pi/k_J=\sigma^2/G\Sigma$, and the characteristic mass is
approximately the square of this half-wavelength times the column
density, $M_J=\sigma^4/G^2\Sigma$. For the velocity dispersion
$\sigma\sim30$ km s$^{-1}$ of the CO gas in the region around the
clusters M4, M5, and M6 (S07; Table 5), and for the average gas mass
column density in the large plateau of CO gas in this region, which is
$\Sigma\sim500\;M_\odot$ pc$^{-2}$, the Jeans mass is
$M_J\sim9\times10^7\;M_\odot$. This is not much different from the
cloud mass in the immediate neighborhood of M4, which is
$5.4\times10^7\;M_\odot$ (for our distance assumption).

Similarly, the wavelength of the instability is $\lambda_J=830$ pc. The
observed separation between M4 and M5 is 4\arcsec~parallel to the major
axis, which is 360 pc at the distance of 18.6 Mpc. Between M5 and M6
the projected distance is also 4\arcsec, but these clusters are
oriented in the direction of the minor axis. Correcting for an
inclination of $40^\circ$, their separation becomes 470 pc. These
cluster separations are about half the current Jeans length.

The cluster separations at the time of their formation, 7 Myr ago,
would have been larger if the clusters have each fallen toward the
center of the large CO plateau in which they are currently located,
which means fallen toward each other. The acceleration rate toward the
center of the plateau is $A=G\Sigma$ for $\Sigma\sim500\;M_\odot$, and
the distance they would have moved in $t=7$ Myr is $0.5At^2=55$ pc.
Thus their separations could have been larger by $\sim110$ pc when they
formed if they each fell toward their common center by 55 pc. This 110
pc, when added to the current separation of $\sim400$ pc, is 60\% of
$\lambda_J$. Considering the uncertainties with $\sigma$, $\Sigma$, the
dispersion relation, and the physical interpretation of $\lambda_J$ for
a complex environment, the overall scale of clustering in this region
is basically consistent with a model in which star formation is driven
by gaseous self-gravity. This conclusion is consistent with the
agreement between the total nuclear star formation rate and the
expectations from the Kennicutt (1998) relation, given the molecular
surface density in the region (S07).

For the average plateau molecular density of $95$ cm$^{-3}$ derived in
Section \ref{sect:env}, the dynamical time for cloud formation is
$\left(G\rho\right)^{-1/2}\sim6.3$ Myr.  If the cloud-forming plateau
moved inward along the bar dustlane at the same speed as the rest of
the gas, 80 km s$^{-1}$, then the formation of the parent clouds that
made the clusters began $\sim13$ Myr ago, 6.3 Myr before the birth of
the clusters, at a distance from the current clusters of $\sim1.0$ kpc.
This is 11'' (deprojected) further out from the current cluster
positions along the bar dustlane.

Figure \ref{schematic_version2} indicates the approximate cloud
formation position. It is at a place in the dustlane where there is
currently an intersection with a filament extending to the interbar
region.  Such filaments would have a timescale for changes comparable
to the timescale for the gas flow from the spiral region to the bar
dust lane. For a pattern speed of 18 km s$^{-1}$ kpc$^{-1}$ and an
orbit speed of 230 km s$^{-1}$ at a mean galactocentric radius of
$\sim3$ kpc (section \ref{sect:where}) the half-orbit time relative to
the bar pattern is $\pi\left(230 \;{\rm km\;s}^{-1}/3\;{\rm
kpc}-18\;{\rm km \;s}^{-1}/{\rm kpc}\right)^{-1}=50$ Myr.  This is
slightly larger than the timescales for cloud and cluster formation
given above. It seems plausible that massive cluster formation was
triggered in the densest part of the bar dustlane by the impact of a
stream of gas flowing from the spiral arm region on the other side of
the galaxy. This is similar to the scenario proposed by B\"oker et al.
(2008) in which cluster formation takes place at the intersection point
between the straight bar dust lane and the circular ILR ring. In our
case, the trigger in the bar dust lane would be the gas falling into
the bar from large radii on the other side.  Our model is closer to
that of Sheth et al. (2000), who noted the presence of dust spurs on
the trailing sides of the bar dust lane in NGC 5383 at approximately
the same positions as HII regions on the leading side. They suggested
that stars formed in the spurs and then pushed through the dust lane to
emerge at the leading side. Asif et al. (2005) made a similar
suggestion for NGC 4151 based on the velocities of HII regions
associated with star formation near the bar dustlane. Zurita \& P\'erez
(2008) thought that this process operated in NGC 1530 based on an age
gradient in HII regions perpendicular to the bar, and they also found
the motion of spur gas toward the dust lanes. The filaments discussed
in the present paper are apparently the same as the spurs in NGC 5383
and NGC 1530. We suggest that star formation may not occur in the spurs
or filaments themselves, but in the intersection points between these
gas streams and the dust lane as a result of local dust lane
compression and triggered instabilities. The emergence of stars or
clusters out the front of the dust lane, if this happens, could then be
the result of the initial stream momentum transferred to the compressed
gas.

Massive dense clusters require very high gas pressures to form. The
total kinematic pressure from stellar motion in the cluster today is
$\sim0.1 GM_{\rm star}^2/R^4\sim10^{11}k_B$ for $M_{\rm
star}\sim10^7\;M_\odot$ and $R\sim5$ pc core radius, and $10^8k_B$ for
$R\sim40$ pc overall. These are 10 to $10^4$ times the average disk gas
pressure discussed in section \ref{sect:env}.  We suggested above that
the inflowing filaments impact the bar dustlane and make the dustlane
pressure high. This is the usual explanation for a dustlane: it is a
shock front in either a spiral arm or a bar caused by the sudden
deflection and compression of incoming gas. The point here is that the
incoming gas is apparently visible in the form of interbar dust
filaments. Because such filaments touch the bar dustlane in a few
discrete points, the pressures should be unusually high there, possibly
triggering star formation. If we consider that the density in a
filament is $\sim0.1$ times that of the dustlane, or $5$ H$_2$
cm$^{-3}$, based on the lower extinction in the filaments shown in
Figure \ref{schematic_version2}, and that their impact speed is the
relative speed between a circular orbit and the bar at this radius,
$214$ km s$^{-1}$ as given in section \ref{sect:where}, then the impact
ram pressure of the filament on the dustlane is $7\times10^7k_B$. This
is about right to generate the dustlane pressure and the pressure in
the cluster-forming cloud. The higher pressure in the cluster core is
presumably the result of self-gravitational contraction in the
molecular cloud.

\section{Conclusions}\label{sect:concl}

The three massive clusters in the center of NGC 1365, along with a
large number of other fainter clusters in the same region, apparently
formed $\sim7$ Myr ago in a giant molecular cloud, the remnants of
which are still visible today (S07). This cloud formed another $\sim6$
Myr earlier in the inner part of the bar dustlane.  The cloud and the
clusters are flowing inward at $\sim80$ km s$^{-1}$ and should soon
join the dust ring inside the ILR as they arc around on the far side of
the nucleus. It is conceivable that subsequent events of star formation
like this will produce a regular sequence of cluster ages around the
inner ring, as observed in M100 by Ryder, Knapen~\& Takamiya (2001) and
Allard et al. (2006), and in several other galaxies by B\"oker et al.
(2008) and Mazzuca et al. (2008). The formation mechanism of the
clusters is probably a gravitational instability in the molecular
cloud, which itself probably formed by self-gravitational gas dynamics
in the moving dustlane, perhaps triggered by the impact of an interbar
filament that is observed at the likely location of cloud formation.

Gas moves in the bar region partly through arching filaments that come
from spiral arms inside corotation, and partly through straight bar
dustlanes, where the gas plunges into the ILR ring.  The accretion rate
from the spirals and the outer parts of the bar dustlanes is
$\sim5.7\;M_\odot$ yr$^{-1}$, while the accretion rate in the inner
parts of the bar dustlanes is $\sim44\;M_\odot$ yr$^{-1}$. The inner
accretion rate, combined with the current gas reservoir, can sustain
the nuclear star formation rate for $\sim0.5$ Gyr.  The bar itself is
probably not much older than this, considering the lack of any source
of gas to replenish the accretion.

Starbursts in the ILR regions of barred galaxies can be driven by rapid
accretion of gas inside corotation as it flows from the spiral region
through the interbar region to the bar and then down the bar dustlanes.
Simulations that suggest a more gradual spiraling of gas to the center
use an equation of state that artificially introduces thermal energy at
rarefaction fronts, providing a source of pressure that is not likely
to be present in a real galaxy.  Massive clusters form because the
self-gravitational pressure in the inner disk can be extremely large,
close to $10^8k_B$.  The most massive clusters may retain their gas
longer than lower-mass clusters because of their higher gravitational
binding energy per unit cluster luminosity. These massive clusters can
then dominate the ionization and emission of [NeII], giving the
impression that they form alone. In fact, there are probably lower-mass
clusters present too, in the usual proportion.

\section{Acknowledgements}

 We are gratefully indebted to F. Bournaud for interesting discussions
 and to the referee for useful comments.  BGE is grateful to
 CEA/Saclay for support during a June 2008 visit when this work
 began. DA thanks CEA/Saclay for supporting a March 2009 trip to the
 Rio observatory, when this work was finalized.

\clearpage

\clearpage

\clearpage

\clearpage

\end{document}